\documentclass[10pt,aps,prd,twocolumn,nofootinbib,superscriptaddress]{revtex4}
\usepackage{amsmath}
\usepackage[utf8]{inputenc}
\usepackage{color}
\usepackage{graphicx}
\usepackage{amssymb}
\usepackage{bm}
\usepackage{acronym}
\usepackage{ifthen}
\usepackage{blindtext}
\usepackage[normalem]{ulem}
\usepackage{hyperref}
\usepackage{etoolbox}
\usepackage{fancyhdr}
\usepackage{xspace}
\usepackage{textcomp}
\usepackage{orcidlink}
\usepackage{multirow}
\usepackage[caption=false]{subfig}
\usepackage{lineno}
\usepackage{tabularx}

\hypersetup{
    colorlinks=true,
    linkcolor=blue,
    filecolor=magenta,      
    urlcolor=black,
		citecolor=red,
		}
\begin{document}

\title{On Nanocones as Gravitational Analog Systems}

\author{F. L. Carneiro\orcidlink{0000-0001-8329-702X}}
\email{fernandolessa45@gmail.com}
\affiliation{Universidade Federal do Norte do Tocantins, 77824-838, Aragua\'ina, TO, Brazil, Brazil}

\author{B. C. C. Carneiro\orcidlink{0000-0002-3791-5007}}
\email{bcccarneiro@gmail.com}
\affiliation{Instituto Federal do Tocantins, 77760-000, Colinas do Tocantins, TO, Brazil} 

\author{D. L. Azevedo\orcidlink{0000-0002-3456-554X}}
\email{david888azv@gmail.com}
\affiliation{Instituto de F\'isica, Universidade de Bras\'ilia, 70910-900, Bras\'ilia, DF, Brazil} 

\author{S. C. Ulhoa\orcidlink{0000-0001-9994-958X}}\email{sc.ulhoa@gmail.com}
\affiliation{Instituto de F\'isica, Universidade de Bras\'ilia, 70910-900, Bras\'ilia, DF, Brazil} \affiliation{Canadian Quantum Research Center,\\ 
204-3002 32 Ave Vernon, BC V1T 2L7  Canada}

%%%%%%%%%%%%%%%%%%%%%%%%%%%%%%%%%%%%%%%%%%%%%%%%%%%%%%%%%%%%%%%%%%%%%%%%%%%%%%%%%%%%%
\begin{abstract}

This study investigates the fundamental properties of graphene and boron nitride (BN) nanostructures, exploring their torsional energy characteristics within the framework of Teleparallel Equivalent of General Relativity (TEGR). By constructing nanocones with disclination defects in these materials, the linear dependence of torsional energy on the disclination angle is analyzed, as predicted by TEGR. The qualitative validation of TEGR's energy expression is supported by simulations, which show a strong correlation between torsional energy and the disclination angle, consistent with theoretical predictions. Additionally, a quantitative analysis is proposed by estimating the coupling constant $\kappa$ associated with TEGR through molecular simulations and Density Functional Theory (DFT) calculations. The results suggest that $\kappa$ reflects the interatomic forces within the materials, providing insights into the nature of spacetime and gravitational interactions on a microscopic scale. These findings contribute to the understanding of material physics and offer implications for the precision and validity of TEGR in describing gravitational phenomena.


\end{abstract}

%%%%%%%%%%%%%%%%%%%%%%%%%%%%%%%%%%%%%%%%%%%%%%%%%%%%%%%%%%%%%%%%%%%%%%%%%%%%%%%%%%%%%
\maketitle

\date{\today}
%%%%%%%%%%%%%%%%%%%%%%%%%%%%%%%%%%%%%%%%%%%%%%%%%%%%%%%%%%%%%%%%%%%%%%%%%%%%%%%%%%%%%
\section{Introduction} \label{sec.1}

Analog models in physics have long existed. A mathematical equivalence between the equations governing certain physical phenomena allows for analogy between them. A good example of this occurred in the realm of electrical circuits. In the early 19th century, when instrumentation for measuring electrical and magnetic quantities was being developed, it was very costly to conduct investigations in circuits containing capacitors and inductors. Therefore, mechanical analog models were employed, which facilitated a better understanding of electrical circuits, as well as simpler experimentation. Analog models thus enable the testing of physical concepts without the need to access the physical system of interest. Experimental black hole physics is still in its infancy and could greatly benefit from the use of gravitational analog models \cite{barcelo}. While experimenting with gravitational waves can shed light on certain aspects of black hole physics, testing other concepts such as Hawking radiation requires the use of analog models, often established in condensed matter. However, there are also acoustic analogs; acoustic black holes, proposed by Unruh \cite{Unruh}, have been used to understand how event horizons can produce a spectrum of thermal radiation \cite{Unruh1}. Hence, gravitational analog models have become important tools for understanding black hole physics.
One such example was the construction of an experimental apparatus where a robophysical model was built to simulate the motion of particles around a Schwarzschild singularity \cite{li2023robophysical}, an experiment that culminated in very satisfactory results compared to the expected values from GR (General Gelativity).

GR is the most widely accepted theory in the literature describing gravitation. However, it is fraught with inconsistencies. Perhaps the most important of these is the definition of gravitational energy. Indeed, there are several attempts to define such an important concept in physics within the scope of general relativity, but none of them meets the necessary requirements for a consistent definition, namely: i) the energy should be independent of the coordinate system used; ii) it should be the component of a covariant four-vector under global Lorentz transformations and therefore should depend on the choice of reference frame. Alternatively, there are approaches that describe the gravitational field, in which the concept of energy is more natural. In this article, an alternative theory dynamically equivalent to general relativity, called Teleparallel Equivalent to General Relativity (TEGR) \cite{maluf}, will be used. In the context of TEGR, there exists a consistent energy-momentum four-vector, so this quantity can be used to experimentally determine which approach correctly describes gravitation. In other words, from a dynamic point of view, both GR and TEGR predict the same behaviors. On the other hand, a consistent definition of energy exists only in TEGR (for a complete discussion of gravitational energy, we encourage the reader to consult Ref. \cite{annalen}). On one hand, there are currently no experiments designed to deal with gravitational energy. On the other hand, the use of gravitational analogue models may lead to clues suggesting the supremacy of TEGR over GR, if gravitational energy is confirmed in such systems.

Graphene is a material predicted in the mid-20th century and synthesized in the early 21st century \cite{grafeno}. It consists of a two-dimensional lattice of carbon atoms arranged in a hexagonal shape, giving graphene highly desired properties such as good conductivity and strength. The equation governing the dynamics of electrons in graphene is relativistic, as the Dirac equation embodies Lorentz transformations. The Dirac equation also naturally couples to the dynamic variables of TEGR, making graphene an excellent analog gravitational system that could allow investigation into the physical reality of gravitational energy. 
	%%%%%%%%%%%%%%%%%%%%%%%%%%%%%%%%%%%%%%%%%%%%%%%
	
Recent studies have explored alternative approaches to describing the electronic and geometric properties of graphene. In particular, metric-based methods have been employed to investigate graphene wormholes and their potential for superconductivity. For instance, the work of Capozziello et al. formulates an effective geometric framework in which superconducting properties emerge from the topology of graphene sheets connected via Chern-Simons bridges \cite{referee1}. Similarly, Ref. \cite{referee2} examines the role of modified gravity-inspired geometric structures in electronic transport within graphene, drawing analogies between crystalline defects and spacetime modifications. These approaches have shown the relevance of curvature-based descriptions in modeling the electronic properties of graphene.

When studying the propagation of fermionic quasiparticles in the presence of a $U(1)$ gauge theory, Zubkov and Volovik \cite{zubkov2015emergent} discovered that, in the presence of elastic deformations, the geometry experienced by the particle was the Weitzenb{\"o}ck geometry. The TEGR is a geometrical gravitational theory within the Weitzenb{\"o}ck geometry. Hence, in this article, we aim to simulate a gravitational field as a non-trivial geometry on a graphene sheet produced by deformations. The connection between geometry and deformations in crystals was established by Katanaev in the Geometric Theory of Defects (GTD) \cite{katanaev2005geometric}, where elastic deformations are characterized by deviations from the metric tensor in Cartesian coordinates, and plastic deformations arise from the emergence of curvature and torsion. Plastic deformations are caused by topological defects, which can manifest as point defects, line defects, or planar defects.
Point defects are often considered in Chern-Simons geometric theory (see, for instance, \cite{katanaev2020point,holz1988geometry}) while planar defects do not seem to hold much significance in analogies between gravitation and defects in crystals. Linear defects can be divided into two categories: disclinations and dislocations. Dislocations are associated with translational deformation, while disclinations are associated with rotational deformations. Disclinations were investigated within the GTD in Ref. \cite{katanaev2021disclinations} and wedge disclinations in Ref. \cite{katanaev2003wedge}. Additionally, the combination of two types of dislocations was considered in Ref.  \cite{katanaev2023combined}. Several applications of the GTD can be found in the literature, such as in the propagation of elastic waves \cite{katanaev2016rotational} and calculating the trajectories of phonons without having to deal with complicated boundary conditions \cite{de1998geodesics}. In this article, we consider a distinct approach: using the notion that plastic deformation on a graphene sheet can be described through the GTD, we employ the tools of TEGR to obtain a behavior for the torsion energy of the defect.

In pursuit of our goal, we consider a graphene sheet as a background flat geometry and impose a gravitational field by inserting a topological defect into the sheet. We choose a disclination defect to construct a well-known nanostructure for graphene, i.e., a nanocone \cite{charlier2001electronic,adhikari2012stabilities,ardeshana2017approach}. Consequently, we evaluate the gravitational energy of the disclination using TEGR and compare the predicted qualitative behavior of the torsional energy with the energy associated with the torsion potential obtained from computational simulations within Universal Force Field (UFF)  . 
The UFF is a versatile molecular mechanics force field used in computational chemistry. It employs empirical parameters to model interatomic interactions, including bonded and non-bonded terms, torsional potentials, and geometric optimizations. UFF is widely applied in molecular simulations for accurate predictions of molecular structures, energies, and properties. The main aspects of this theory necessary for the comprehension of the simulations will be presented in this article.
We also infer about the possibility of the physical origin of the coupling constant in graphene and obtain a numerical value for the energy within the order of magnitude given by the simulations. Therefore, we aim to address these aspects in this article to support future experiments designed to measure gravitational energy, especially since the synthesis of nanocones is already a reality \cite{experimental}.

%%%%%%%%%%%%%%%%%
Thus, this article is divided as follows: In Section \ref{sec.2}, we briefly present the concepts of gravitational energy and momentum within the scope of TEGR. In section \ref{sec.2.1} we introduce the disclination topological defect considered. In Section \ref{sec.3}, we present the general aspects of the UFF. In Section \ref{sec.4}, we present the results and respective analyses. Finally, in the last section, we present our conclusions.

%%%%%%%%%%%%%%%%%%%%%%%%%%%%%%%%%%%%%%%%%%%%%%
\section{Teleparallelism Equivalent to General Relativity (TEGR)} \label{sec.2}
%%%%%%%%%%%%%%%%%%%%%%%%%%%%%%%%%%%%%%%%%%%%%%

TEGR is a geometrical theory of gravity that was first considered by Einstein in an attempt to unify gravity and electromagnetism. Several decades after being abandoned by Einstein, M{\o}ller revived the idea with another objective: to establish a true energy-momentum for the gravitational field \cite{moller1964conservation}. Despite the many decades of GR, researchers have not been able to define a true energy-momentum tensor for the gravitational field, with several pseudo-tensors defined in the literature, all of which depend on the choice of coordinates. Therefore, a definition for the energy of a gravitational field is only possible globally, i.e., by integrating the pseudo-tensors across the entire spacetime. Such integration is possible only if the spacetime is asymptotically flat at spatial infinity. After a few years of attempting to define a true energy-momentum tensor for the gravitational field, M{\o}ller realized that the main problem with GR was the fact that its fundamental variable is the metric tensor $g_{\mu\nu}$ and concluded that he could solve the problem by formulating GR with the tetrad field $e_{a\mu}$ as its fundamental variable. Furthermore, Maluf realized that the field equations of the theory allow for a definition of a true energy-momentum tensor for the gravitational field. In the following, we follow the development of TEGR as presented in Ref. \cite{maluf}.

The tetrads $e_{a\mu}$ have 16 independent components, with one $\mu$ (Greek) raised and lowered by the metric tensor of spacetime $g_{\mu\nu}$ and $a$ (Latin) raised and lowered by the metric tensor $\eta_{ab}$ of the tangent space (fiber bundle) to the spacetime at the event $x^{\mu}$. Thus, the tetrads project quantities $v^\mu$ from spacetime into $v^a$ quantities of the tangent space, i.e., $v^a=e^{a}\,_\mu v^\mu$. Given the properties of lowering and raising indices by the metric tensor, the metric tensor itself can be obtained by the relation
\begin{equation}\label{eq1}
	g_{\mu \nu}= e_{a \mu} e^{a}\,_{\nu}\,.
\end{equation}
In view of equation (\ref{eq1}), the tetrads are more fundamental than the metric tensor. The tetrads $e_{a\mu}$ transform as covariant 4-vectors under coordinate transformations in the Greek index and under the $SO(3,1)$ group in the Latin index.

The geometry in which TEGR is established is the Weitzenb\"{o}ck geometry. Given the prominence in the literature of topological defects considered within Riemann-Cartan geometry, we depart from Riemann-Cartan to the Weitzenb\"{o}ck one, where the Levi-Civita connection is responsible for the rotation symmetry instead of the affine connection in Riemann-Cartan geometry. Therefore, we shall show that TEGR admits linear rotations (disclinations) and translations (dislocations) as topological defects in its geometry, similar to Riemann-Cartan.

In a manifold endowed with metricity, the structure equations can be obtained from the co-frame 1-form $e^a=e^{a}\,_\mu dx^\mu$ and the connection 1-form $\omega^{a}\,_{b}=\omega^{a}\,_{b\mu}dx^{\mu}$, where $\omega^{a}\,_{b\mu}$ are the 0-form components of the connection. The torsion 2-form $T^a$ of the manifold can be obtained by the exterior total derivative $D$ of the co-vector, i.e.,
\begin{align}
	T^a = D\, e^a &= d\, e^a + \omega^{a}\,_{b}\wedge e^b\nonumber\\
	&=\frac{1}{2}T^{a}\,_{\mu\nu}dx^{\mu}\wedge dx^{\nu}
\end{align}
where $d$ indicates the exterior derivative, $\wedge$ denotes the wedge product, and
\begin{equation}
	T^{a}\,_{\mu\nu}=\partial_{\mu}e^a\,_{\nu}-\partial_{\nu}e^a\,_{\mu}+\omega^{a}\,_{b\mu}e^{b}\,_{\nu}-\omega^{a}\,_{b\nu}e^{b}\,_{\mu}
\end{equation}
are the  0-form components of the torsion tensor. The curvature 2-form is
\begin{align}
	R^{a}\,_{b}&=d\omega^{a}\,_{b} + \omega^{a}\,_{c}\wedge \omega^{c}\,_{b}\nonumber\\
	&=\frac{1}{2}R^{a}\,_{b\mu\nu}dx^{\mu}\wedge dx^{\nu}\,,\label{eq4}
\end{align}
where
\begin{align}
	R^{a}\,_{b\mu\nu} &= \partial_{\mu}\omega^{a}\,_{b\nu} - \partial_{\nu}\omega^{a}\,_{b\mu}\nonumber\\
	& + \omega^{a}\,_{c\mu}\omega^{c}\,_{b\nu} - \omega^{a}\,_{c\nu}\omega^{c}\,_{b\mu}
\end{align}
are the components 0-form of the curvature tensor.

The geometry of TEGR may be understood as a special case of Cartan geometry, i.e., when the curvature tensor (\ref{eq4}) constructed from the affine connection vanishes. In this case, the affine connection $\omega^{a}\,_{c}$ plays no role in the dynamics of the field \cite{maluf}. Hence, we may consider it to be zero. Given the relation of the curvature tensor with rotational symmetries, it is possible to construct a geometry in which the parallel transport of a vector along a closed curve occurs, i.e., a parallelism at a distance ($\tau\eta\lambda\epsilon$ =``tele" = far). Thus, in such a geometry, the tetrad has zero covariant derivative, i.e.,
\begin{equation}\label{eq6}
	\nabla_{\mu}e_{a\nu}=\partial_{\mu}e_{a\nu}-\Gamma^{\lambda}\,_{\mu\nu}e_{a \lambda}=0\,.
\end{equation}
The connection that guarantees distant parallelism is given by
\begin{equation}
	\Gamma^{\lambda}\,_{\mu\nu}=e^{a\lambda}\partial_{\mu}e_{a\nu}
\end{equation}
whose torsion tensor components are given by
\begin{equation}
T^{a}\,_{\mu\nu}=\partial_{\mu} e^{a}\,_{\nu}-\partial_{\nu}e^{a}\,_{\mu}\,.\end{equation}
From the torsion tensor components, the contortion tensor
\begin{eqnarray}
K_{\lambda\mu\nu}&=&\frac{1}{2}(T_{\lambda\mu\nu}+T_{\mu\lambda\nu}-T_{\nu\lambda\mu})\,,\label{3}
\end{eqnarray}
can be evaluated. Since the affine connection is chosen to be zero, the components of the Levi-Civita connection $\mathring{\omega}_{ab\mu}$ can be evaluated using the relation
\begin{equation}\label{eq10}
\omega_{ab\mu}= \mathring{\omega}_{ab\mu}+ K_{ab\mu}=0
\end{equation}
for the contortion tensor components.

After establishing the geometry on which TEGR is constructed, we can define the theory on this geometry. First, we define the superpotential as
\begin{equation}
\Sigma^{abc}=\frac{1}{4} (T^{abc}+T^{bac}-T^{cab}) +\frac{1}{2}(
\eta^{ac}T^b-\eta^{ab}T^c)\,,
\end{equation}
where $T^\mu=T^{\nu}\,_{\nu}\,^{\mu}$.
Contracting the superpotential with the torsion, we may construct the Weitzenb\"{o}ck invariant $T=\Sigma^{abc}T_{abc}$ and the Lagrangian density of TEGR
\begin{eqnarray}
\mathfrak{L}(e_{a\mu})&=& -\kappa\,e \Sigma^{abc}T_{abc} -\frac{1}{c}\mathfrak{L}_M\nonumber\\
&=&-\kappa\,e\,(\frac{1}{4}T^{abc}T_{abc}+
\frac{1}{2} T^{abc}T_{bac} -T^aT_a) \nonumber\\
&-&\frac{1}{c}\mathfrak{L}_M\,,\label{eq12}
\end{eqnarray}
where $e=det(e^{a}\,_{\mu})$, $\kappa=c^{3}/16\pi G$ in SI units, and $\mathfrak{L}_M$ stands for the matter-radiation fields Lagrangian density.
The torsion invariant, and hence the Lagrangian density, is equal to the curvature scalar constructed from the Levi-Civita connection plus a total divergence term, i.e.,
\begin{align}
eR(e)&= -e(\frac{1}{4}T^{abc}T_{abc}+\frac{1}{2}T^{abc}T_{bac}-T^aT_a)\nonumber\\
&+2\partial_\mu(eT^\mu)\,.\label{eq13}
\end{align}
The Lagrangian density (\ref{eq12}) is invariant under coordinate transformations and under global Lorentz transformations, but not under local Lorentz transformations. The total divergence term $2\partial_\mu(eT^\mu)$ makes (\ref{eq13}) invariant under local Lorentz transformations, and its inclusion in the Lagrangian density would make it invariant under such transformations. In this case, we would have another theory than TEGR. However, the equations of motion would not be altered.

The dynamics of the theory can be obtained by varying the Lagrangian density (\ref{eq12}) with respect to $e_{a\mu}$, yielding
\begin{equation}\label{eq14}
\partial_\nu\left(e\Sigma^{a\lambda\nu}\right)=\frac{1}{4\kappa}
e\, e^a\,_\mu( t^{\lambda \mu} + T^{\lambda \mu})\,,
\end{equation}
where $T^{\lambda\mu}$ is energy-momentum tensor of the matter-radiation field and
\begin{equation}\label{eq15}
t^{\lambda \mu}=\kappa\left[4\,\Sigma^{bc\lambda}T_{bc}\,^\mu- g^{\lambda
\mu}\, \Sigma^{abc}T_{abc}\right]\,.
\end{equation}
Although the Lagrangian density (\ref{eq12}) is not invariant under local Lorentz transformations, the field equations (\ref{eq14}) are covariant under coordinate transformations and under the full group of Lorentz transformations. It can be shown that (\ref{eq14}) may be written as $R_{a\mu}-\frac{1}{2}e_{a\mu}R=\frac{1}{2c\kappa}T_{a\mu}$, i.e., they are equivalent to Einstein's equations. Therefore, TEGR yields the same dynamics as GR but allows for the definition of a true energy-momentum tensor.

The energy-momentum tensor can be obtained directly from (\ref{eq14}). By noticing that the superpotential is antisymmetric in the last two indices, we have
\begin{equation}\label{eq16}
\partial_\lambda\partial_\nu\left(e\Sigma^{a\lambda\nu}\right)=0\,.
\end{equation}
Integrating (\ref{eq16}) over an arbitrary three-volume $V$ and separating the time and spatial coordinates, we arrive at a continuity (or balance) equation
\small
\begin{equation}\label{eq17}
\frac{d}{dt}\int_V d^3x \,e\,e^a\,_\mu(t^{0\mu}+ T^{0\mu})=-\oint_{\partial V} dS_{j}\,\,e\,e^a\,_\mu(t^{j\mu}+ T^{j\mu})\,,
\end{equation}
\normalsize
where $dS_{j}$ is the infinitesimal area element of $\partial V$ oriented along the $j$ direction. If the RHS of (\ref{eq17}) is taken at spatial infinity, we have a conservation equation for the quantity
\begin{equation}
P^a = \int_V d^3x \,e\,e^a\,_\mu(t^{0\mu}+ T^{0\mu})\,,
\end{equation}
and we may identify (\ref{eq15}) as the energy-momentum tensor of the gravitational field. The tensor (\ref{eq15}) is a true tensor, i.e., it transforms covariantly under coordinate transformations. Given this identification, the quantity
\begin{equation}\label{eq19}
P^a =4\kappa\, \int_V d^3x \,\partial_\nu\left(e\,\Sigma^{a0\nu}\right)
\end{equation}
is the total momentum 4-vector of the gravitational field contained within $V$. The 4-momentum (\ref{eq19}) is invariant under coordinate transformations of the three-dimensional space, under time reparametrizations, and covariant under global Lorentz transformations.

%%%%%%%%%%%%%%%%%%%%%%%%%%%%%%%%%%%%%%%%%%%%%%
\section{Topological Defects in spacetime} \label{sec.2.1}
%%%%%%%%%%%%%%%%%%%%%%%%%%%%%%%%%%%%%%%%%%%%%%

Linear topological defects in spacetime can be constructed using the Volterra process, which distorts spacetime. This process involves cutting and soldering forms that allow a section of spacetime to be removed and the edges glued together. The details of this process can be found in Ref. \cite{puntigam1997volterra}. Ten defects can be constructed using this method: four temporal ones, three spatial dislocations, and three spatial disclinations.

A dislocation is characterized by the Burgers vector $b^{(i)}$ oriented along the $(i)$ directions, representing the translational deficit in the topology. The Burgers vector can be evaluated from the torsion tensor using the relation \cite{katanaev2005geometric}
\begin{equation}\label{eq20}
	b^{(i)}=\frac{1}{2}\int_{S}T^{(i)}\,_{\mu\nu} dx^{\mu} \wedge dx^{\nu}\,,
\end{equation}
where $S$ is the surface orthogonal to the defect. Hence, from (\ref{eq20}), we can see that the torsion tensor represents the density of dislocations in space. If the integral (\ref{eq20}) along a surface $S$ oriented by $dx^{\mu} \wedge dx^{\nu}$ evaluates to a nonzero value, we can conclude that space has a dislocation defect. Applying Stokes' theorem, we can write (\ref{eq20}) as
\begin{equation}\label{eq21}
	b^{(i)}=\oint_{\partial S}e^{(i)}\,_{\mu} \, dx^{\mu}\,,
\end{equation}
i.e., the defect is ``measured" by an observer carried along the boundary of $S$ as the closing failure of the closed Burgers circuit $\partial S$. This result is expected, as torsion is associated with translational symmetry. Therefore, given that the Weitzenb\"{o}ck geometry allows a nonzero torsion tensor, it is expected that TEGR allows for the characterization of dislocation defects.

A disclination defect is associated with angular deficits in spacetime and can be evaluated from the matrix \cite{katanaev2005geometric}
\begin{equation}\label{eq22}
	\Omega^{(i)(j)}=\frac{1}{2}\int_{S}R^{(i)(j)}\,_{\mu\nu}\, dx^{\mu} \wedge dx^{\nu}\,.
\end{equation}
The orientation of the angle deficit axis can be obtained by the path-ordered exponential of  $\Omega^{(i)(j)}$, i.e.,
\begin{equation}\label{eq23}
\mathcal{G}^{(i)(j)}=\exp{(2\pi\,\Omega^{(i)(j)})}\,.
\end{equation}
The matrix $\mathcal{G}^{(i)(j)}$, called Frank matrix, represents the rotation matrix along the axis containing the angle deficit, i.e., the rotation angle represents the amount that we must rotate to perform a full $2\pi$ lap in spacetime.
The curvature tensor of the affine connection is zero for the Weitzenböck geometry. However, the curvature tensor of the Levi-Civita connection $\mathring{\omega}_{ab\mu}$ is nonzero in the general case. This curvature tensor allows for a consistent definition of disclination in TEGR. This feature will be demonstrated here for the defect considered.

A conical defect along the $z$ direction is characterized by the disclination angle parameter $\Delta\phi = 2\pi\big(1-\beta(r)\big)$ in the line element
\begin{equation}\label{eq24}
	ds^{2}=-dt^{2}+dr^{2}+\beta(r)^{2}r^{2}d\phi^{2}+dz^{2}\,.
\end{equation}
When $\beta=1$, we recover Minkowski spacetime. The disclination angle can be evaluated from (\ref{eq24}) by computing the length of the circumference along the $z$ axis, but we will consider relations (\ref{eq22}) and (\ref{eq23}). First, we need a set of tetrads associated with the metric (\ref{eq24}) and adapted to a static observer. One set satisfying the condition is given by
\begin{equation}\label{eq25}
e_{a\mu}=\left(
\begin{array}{cccc}
 -1 & 0 & 0 & 0 \\
 0 & \cos \phi  & -r \sin \phi \, \beta (r) & 0 \\
 0 & \sin \phi  & r \cos \phi \, \beta (r) & 0 \\
 0 & 0 & 0 & 1 \\
\end{array}
\right)\,.
\end{equation}
From tetrads (\ref{eq25}), we evaluate the only non-zero component of $T^{\lambda\mu\nu}$ as
\begin{equation}
	T^{221}=\frac{1-\partial_{r}(r\,\beta)}{r^3\beta^3}\,,
\end{equation}
and also the only non-zero component of the Levi-Civita connection as
\begin{equation}\label{eq27}
	\mathring{\omega}^{(1)(2)}\,_{2}=1-\partial_{r}(r\,\beta)\,.
\end{equation}
From (\ref{eq27}), the only non-zero component of $R^{ab}\,_{\mu\nu}$ is given by
\begin{equation}
	R^{(1)(2)}\,_{21}=2\,\partial_{r}\beta + r\,\partial^{2}_{r}\beta\,.
\end{equation}
We can see that the curvature tensor is non-zero. The Frank matrix can be evaluated from (\ref{eq22}) directly, or we may notice that if the rotations are contained in a spatial plane orthogonal to the defect, as is the case considered here, we have the subgroup $SO(2)$ where the quadratic components of the connection vanish, and we may write (\ref{eq22}) as
\begin{align}\label{eq29}
	\Omega^{(i)(j)}&=\frac{1}{2}\int_{S} \Big( \partial_{\mu}\mathring{\omega}^{(i)(j)}\,_{\nu}  - \partial_{\nu}\mathring{\omega}^{(i)(j)}\,_\mu \Big) \, dx^{\nu} \wedge dx^{\mu}\nonumber\\
	&= \oint_{\partial S}\mathring{\omega}^{(i)(j)}\,_{\mu} \, dx^{\mu}\,.
\end{align}
Hence,
\begin{equation}
	\Omega^{(1)(2)}=\oint_{\partial S} \Big( 1-\partial_{r}(r\,\beta) \Big)  \, d\phi\,.
\end{equation}

The evaluation of (\ref{eq29}) involves modeling the defect core, i.e., choosing a function $\beta(r)$. A simple model can be constructed by considering two regions: (i) a disclination core, where we have a flat spacetime, and (ii) an outside region, where we also have a flat spacetime, i.e., both regions are locally flat. In this case, we may express the function $\beta (r)$ as
\begin{equation}\label{eq31}
\beta(r)=
\begin{cases} 
      \beta_{0}\,, & r>a\,, \\
      1\,, & r\leq a\,,
\end{cases}
\end{equation}
where $a$ is the dislocation core radius. We can see that the curvature tensor will be zero everywhere but at $r=a$. Therefore, the spacetime is locally flat but not globally. By choosing a circuit $\partial S$ with radius $R>a$, we have
\begin{align}
	\mathcal{G}^{(i)(j)}&=exp\left[\left(\begin{array}{cc}
 0 & 2\pi(1-\beta_{0}) \\
   -2\pi(1-\beta_{0}) & 0  \\
\end{array} \right)\right]\nonumber\\
&=\left(\begin{array}{cc}
 \cos(2\pi(1-\beta_{0})) & \sin(2\pi(1-\beta_{0}))  \\
 -\sin(2\pi(1-\beta_{0}))& \cos(2\pi(1-\beta_{0})) \\
\end{array} \right)\,.
\end{align}
i.e., the rotation matrix of the disclination angle $\Delta\phi=2\pi(1-\beta_{0})$ along the $z$ axis.
Thus, TEGR allows torsion and curvature and is consistent in computing topological defects in spacetime, with the rotational defects given by the Levi-Civita connection constructed out of the contortion tensor given by (\ref{eq10}).

The 4-momentum (\ref{eq19}) of (\ref{eq25}) has already been obtained in the literature \cite{maluf1997gravitational,maluf2001space,carneiro2020quantization}. However, from the torsion (\ref{eq22}), it is straightforward to obtain the only non-zero component of $\Sigma^{a0\nu}$ as
\begin{equation}
	\Sigma^{(0)01}=\frac{1-\partial_{r}(r\beta)}{2r\beta}\,.
\end{equation}
Hence, the zeroth component of $P^{a}$ contained within a cylinder of radius $R>a$ and length $L$ is given by
\begin{align}
	P^{(0)}&=2\kappa \int_{V}\partial_{r} \big( 1-\partial_{r}(r\beta)\big)dV\nonumber\\
	&=2\kappa\int_{0}^{L} dz\int_{0}^{2\pi}d\phi \, \big( 1-\partial_{r}(r\beta)\big)\Big|_{r=R}\nonumber\\
	&=4\pi\, \kappa\, L (1-\beta_{0})\nonumber\\
	&=\frac{c^{3}}{8\pi G} \, L \, \Delta\phi\,.\label{eq34}
\end{align}
From (\ref{eq34}), we see that with TEGR, we expect the energy $E=c P^{(0)}$ to not depend on the radius $R$ and, more importantly, to be linear with the angle deficit $\Delta{\phi}$.

Our goal is to construct a graphene monolayer that has the topological structure of (\ref{eq24}), calculate the energy of the structure associated with its molecular torsion potential, and determine the qualitative validity of (\ref{eq34}) and TEGR itself.

%%%%%%%%%%%%%%%%%%%%%%%%%%%%%%%%%%%%%%%%%%%%%%
\section{Universal Force Field in Molecular Mechanics} \label{sec.3}
%%%%%%%%%%%%%%%%%%%%%%%%%%%%%%%%%%%%%%%%%%%%%%

The complexity of multi-electron atoms makes it challenging to apply quantum mechanics directly to solve their problems. In such scenarios, we must resort to approximate methods alongside the many-electron wave function trial $\psi \left (\vec{r_1}, \vec{r_2}, \vec{r_3},... , \vec{r_n}\right)$ \cite{bretonnet2017basics,viana2004quantitative} to solve the appropriated Schr\"odinger equation, and therein obtain the behavior of these systems. Due to the computational limitations, molecular mechanics, a classical approach, becomes indispensable for large atomic systems, aiding in understanding molecular properties and behavior in diverse settings. At the core of molecular mechanics lies the UFF, a framework crucial for simulating molecular interactions \cite{rappe1992uff,jaillet2017uff}.  The UFF is established empirically, parameterizing potential energy functions based on experimental data \cite{mackerell2004empirical,bradley1972ap}.

\begin{figure}[htbp]
\centering
\includegraphics[width=0.52\textwidth]{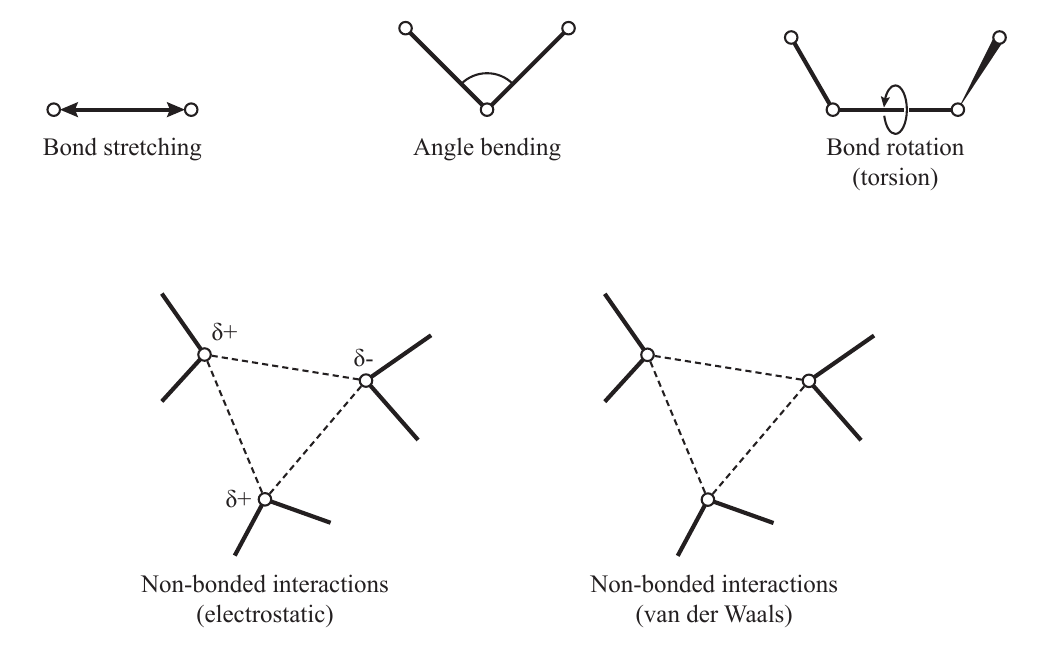}
\caption{A geometrical representation of some interactions described by the UFF potential is presented, highlighting some  different types of interactions, such as the stretching, bending and twisting of atomic bonds, and non-bonded interactions.}
\label{fig0}
\end{figure}

It makes things easier by only looking at the positions of the nuclei and ignoring the movements of the electrons. It then figures out the effective potential by adding up the potentials of all the atoms, which means it treats the atoms like particles. A schematic representation illustrating each type of interaction is shown in Figure \ref{fig0}

%%%%%%%%%%%%%%%%%%%%%%%%%%%%%%%%%%%%%%%%%%%%%%%%
The development of UFF uses a combination of experimental data and high-level quantum mechanical calculations to parameterize the force field for different types of atoms and their interactions \cite{mayne2013rapid,vanommeslaeghe2014molecular}. The primary objective of UFF is to achieve molecular equilibrium geometries and ascertain material properties at a reduced computational cost.  
The versatility of UFF makes it suitable for a wide range of applications in molecular modeling.  This has significant implications for fields such as materials science, biochemistry, and pharmacology \cite{mackerell2004empirical}.
For example, in drug design, UFF can be used to model the interaction between a drug molecule and its target protein, providing information on the binding affinities and conformational changes involved \cite{tute1995drug}. In materials science, UFF facilitates the investigation of molecular structures and properties of polymers, nanomaterials, and complex alloys. By simply modeling the forces between atoms through an analytical form of potentials with some parameters, researchers not only can predict mechanical properties, stability, and reactivity but could guide the design of new materials with customized properties. The greatest strength of UFF is the wide applicability and parametrization for almost all elements. This universality makes it a valuable tool for researchers working with diverse chemical systems without the need to switch between different force fields. \\ 
In this methodology,  the total potential energy of the system is represented as the superposition of multiple individual potential interactions, the general form utilized in obtaining the findings of this paper is shown at~(\ref{potential}).  Where the individual contributions to $V(R)$ describes the following components types in order: bond stretching, represented by a Morse potential; bond angle deformations; torsional (dihedral) interactions; improper (out-of-plane bending) interactions; couplings between distinct chemical bonds; couplings between bond angles; couplings between bond angles and bond lengths; interactions between torsional angles and bond angles; couplings between improper angles; the van der Waals potential (Lennard-Jones terms); and the electrostatic interactions between atomic pairs.  The final double summation in ~(\ref{potential}) refers to non-bonded interactions.

\begin{widetext}
\begin{equation}
\begin{aligned}
    V(R) &= \sum_b D_b \left[ 1 - e^{-a(b - b_0)} \right]^2 
    + \sum_\theta H_\theta (\theta - \theta_0)^2 
    + \sum_\phi H_\phi \left[ 1 - s\cos(n\phi) \right] \\
    &\quad + \sum_\chi H_\chi \chi^2
    + \sum_b \sum_{b'} F_{b b'} (b - b_0)(b' - b_0') 
    + \sum_\theta \sum_{\theta'} F_{\theta \theta'} (\theta - \theta_0)(\theta' - \theta_0') \\
    &\quad + \sum_b \sum_\theta F_{b\theta} (b - b_0)(\theta - \theta_0) 
    + \sum_\phi F_{\phi \theta \theta'} \cos \phi (\theta - \theta_0)(\theta' - \theta_0') \\
    &\quad + \sum_\chi \sum_{\chi'} F_{\chi \chi'} \chi \chi' 
    + \sum_i \sum_{j > i} \left[ \frac{A_{ij}}{r_{ij}^{12}} - \frac{B_{ij}}{r_{ij}^6} + \frac{q_i q_j}{r_{ij} \varepsilon} \right].
\end{aligned}
\label{potential}
\end{equation}
\end{widetext}
In equation (\ref{potential}), $ b $ represents the bond length between two atoms in a molecule, $ \theta $ is the bond angle formed between three consecutive atoms, and $ \phi $ is the torsional angle between four consecutively bonded atoms, defining the rotation about a bond. The term $ \chi $ corresponds to the out-of-plane angle, which quantifies deviations from planarity in structures such as aromatic rings and carbonyl groups. The parameter $ a $ controls the stiffness of the bond stretching energy in the Morse potential, while $ \epsilon $ represents either the depth of the potential well in the Lennard-Jones interaction or the permittivity of the medium in the electrostatic term, depending on the context. The subscript $ 0 $ denotes equilibrium or reference values of structural parameters. Additionally, $ b $ and $ b' $ refer to different bond lengths, and $ V $ and $ V' $ indicate distinct potential energy surfaces.

The last term in equation (\ref{potential}) accounts for non-bonded interactions, considering all pairs of atoms (\textit{i} and \textit{j}) in the molecule and providing an energetic treatment of atomic interactions. In this term, $r_{ij} $ represents the distance between two interacting particles (nuclei), $q_i $ and $ q_j $ denote the atomic charges, $\epsilon$ is the depth of the potential well (Lennard-Jones energy minimum), and $ \sigma_{ij} $ is the characteristic distance at which the Lennard-Jones potential equals zero. This term encapsulates electrostatic interactions (Coulombic) and van der Waals forces (Lennard-Jones), among others, contributing to the overall energy of the molecular system \cite{leach2001molecular}.

The parameters $ D_b $, $ H_\theta $, $ H_\phi $, and $ H_\chi $ determine the energetic contributions of different molecular deformations. The term $ D_b $ represents the bond dissociation energy and appears in the Morse potential, describing the energy required to break a chemical bond. The constants $ H_\theta $ and $ H_\phi $ correspond to the force constants for bond angle bending and torsional rotation, respectively, governing the resistance of the molecule to angular deformations. The parameter $ H_\chi $ is the out-of-plane bending force constant, which measures the resistance of planar structures, such as aromatic rings and carbonyl groups, to deviations from planarity.  
The terms $ F_{b b'} $, $ F_{\theta \theta'} $, $ F_{b\theta} $, and $ F_{\phi \theta \theta'} $ represent cross-interaction parameters that account for couplings between different molecular distortions. Specifically, $ F_{b b'} $ and $ F_{\theta \theta'} $ describe interactions between pairs of bond lengths and bond angles, respectively, while $ F_{b\theta} $ captures the coupling between bond stretching and angle bending. The term $ F_{\phi \theta \theta'} $ quantifies the influence of bond torsion on bond angle variations.  
Finally, the coefficients $ A_{ij} $ and $ B_{ij} $ appear in the Lennard-Jones potential, which describes van der Waals interactions between non-bonded atoms. The parameter $ A_{ij} $ governs the strength of the repulsive interaction, typically associated with Pauli exclusion effects, while $ B_{ij} $ determines the attractive interaction due to induced dipole-dipole forces.
All these parameters are adjusted empirically or derived from quantum mechanical calculations to accurately reproduce the mechanical and structural properties of molecules.

Given the significance of energy torsion in this paper, we will subsequently emphasize just the terms that denote the torsional components of the potential energy. They are
\begin{align}
    V_{torsion}&=\sum_\phi H_\phi \left[ 1 - s\cos(n\phi) \right]\nonumber \\  
    &+\sum_\phi F_{\phi \theta \theta'} \cos \phi (\theta - \theta_0)(\theta' - \theta_0')\nonumber \\      
		&+\sum_\chi H_\chi \chi^2 +\sum_\chi \sum_{\chi'} F_{\chi \chi'} \chi \chi'  \label{torsion}\,.
\end{align}

The rotation around a chemical bond; the coupling between dihedral angles and bond angles models how torsional motion interacts with bond bending; improper torsions (out-of-plane bending) account for deformations in planar systems such as aromatic rings and carbonyl groups; the coupling between different improper torsions represents interactions between multiple out-of-plane distortions. With that in mind, we conducted molecular mechanics calculations using the Forcite module of Materials Studio\cite{rappe1992uff}. To achieve the equilibrium geometry, the program employs an iterative approach designed to attain the global minimum of the potential energy surface, signifying a state where the forces acting on the system are balanced to zero.  This self-consistency cycle continues until the convergence conditions are satisfied, usually following several iterations. To assure precision, we employed an integration grid to sample the potential energy landscape during geometric optimizations.  Geometric optimization entails the iterative modification of atomic coordinates to identify the configuration that represents the minimal energy state. Our convergence criteria encompassed restricting the maximum atomic displacement to $5.0\times10^{-3} \AA$, the maximum energy variation to $4.34 \times 10^{-7} eV$, the maximum force to $2.7 \times10^{-2} eV/\AA$ , and the maximum stress to $0.01\,GPa$.  The tight criteria were crucial for obtaining accurate and dependable findings in our computations, guaranteeing that the resulting molecular configurations were energetically advantageous and physically significant.  Subsequently, we calculated the aggregate of torsional energy components to evaluate the energy required for deforming the graphene surface on nanocones.

%%%%%%%%%%%%%%%%%%%%%%%%%%%%%%%%%%%%%%%%%%%%%%
\section{Results and analysis} \label{sec.4}
%%%%%%%%%%%%%%%%%%%%%%%%%%%%%%%%%%%%%%%%%%%%%%

The first step in corroborating result (\ref{eq34}) is to choose a material that easily emulates the three-dimensional space section of spacetime. The Volterra process can be performed on a nanoscale in a graphene sheet. Graphene is a very stable material with strong carbon-carbon bonds  (Young's modulus of $1TPa$) that can accommodate disclination defects with minimal local deformations for specific disclination angles. This feature is important for properly describing the material by the line element (\ref{eq24}), where, for function (\ref{eq31}), there must be no local deformations, i.e., the resulting geometry must be locally flat.

The total energy of the structure has several contributions, including those related to (\ref{torsion}). Part of this energy, i.e., the one associated with the torsion potential, has geometric origins and it is global, meaning it is linked to the overall shape of the structure. This energy is caused by the disclination defect and is not expected to be significantly affected by other atomic phenomena, such as the number of atoms in the structure. Given that the potential from the UFF is empirical, we can regard its results as quasi-experimental. Our goal is to identify this ``energy derived from experimental data" with the torsion energy described by TEGR.

In this section, we consider disclination defects in graphene sheets in subsection \ref{sec.4.1} and in boron nitride sheets, graphene-like sheets with boron and nitrogen atoms alternating positions, in subsection \ref{sec.4.2}. Energy will be measured in electron volts (eV) and distances in angstroms ($\AA$, $10^{-10}$ meters) unless otherwise stated.

%%%%%%%%%%%%%%%%%%%%%%%%%%%%%%%%%%%%%%%%%%%%%%
\subsection{Graphene nanocones} \label{sec.4.1}

A graphene sheet is not a strictly two-dimensional material, as it has a small but non-zero thickness. This thickness allows the description of the metric tensor of a non-deformed sheet using the metric tensor of the three-dimensional $t=\text{constant}$ hypersurface of Minkowski spacetime.

The sheets are composed of carbon hexagons with partial single-double bonds (aromatic rings). This hexagonal structure of the sheet allows it to accommodate global disclinations only with specific discrete values given by integer multiples of $60^{\circ}$, i.e.,
\begin{equation}\label{angles}
	\Delta \phi = n \frac{\pi}{3}\,,
\end{equation}
where $n = 1, 2, 3, 4, 5$. A graphene sheet with a disclination topological defect characterized by a disclination angle given by (\ref{angles}) is called a nanocone. It has been found that nanocones of silicon carbide (with a similar structure) are stable only for $n = 1, 2, 5$ \cite{alfieri2011structural}. However, by constructing pentagons at the top of the silicon carbide nanocone, dangling bonds are avoided, and optimized structures can also be constructed for $ n = 3, 4$ \cite{adhikari2012stabilities}. Therefore, to accommodate as much data as possible, we shall consider structures with a flat top containing one or more carbon pentagons, as silicon carbide is also hexagonal. This approach may make the disclination core less precise in describing the actual defect, but it is necessary to achieve stable structures and avoid local bending. Consequently, we expect a minor error source due to the deviation of the disclination core from a perfect circle described by (\ref{eq31}).

The second step is to construct the Volterra process for the graphene sheet. We consider distinct sheets of various radii that are nearly circular. Given the hexagonal symmetry, it is impossible to achieve a perfectly circular sheet, so we anticipate another minor error due to the deviation from a perfect circle of radius $ R $ in the integration contained in (\ref{eq34}).

\begin{figure}[htbp]
	\centering
		\includegraphics[width=0.40\textwidth]{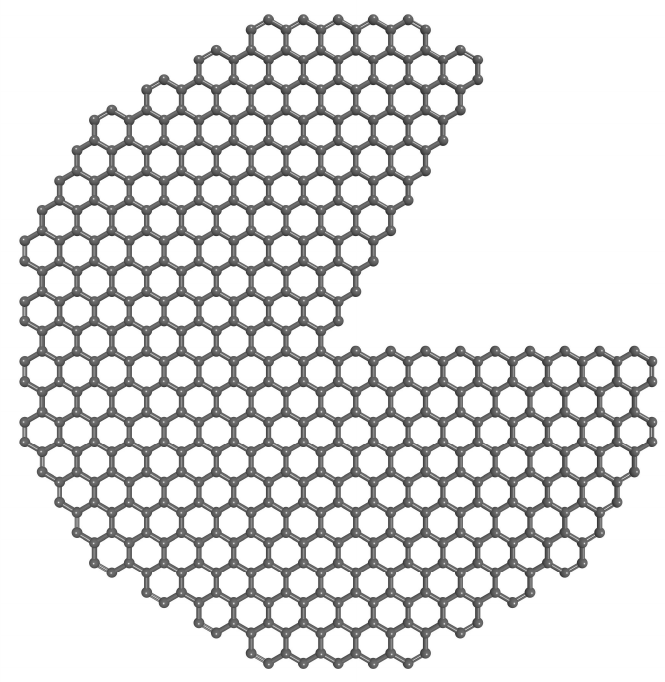}
	\caption{Graphical representation of the removal of a section $\Delta\phi=60^{\circ}$ from graphene Sheet 1.}
	\label{fig1}
\end{figure}

\begin{figure}[htbp]
	\centering
		\includegraphics[width=0.40\textwidth]{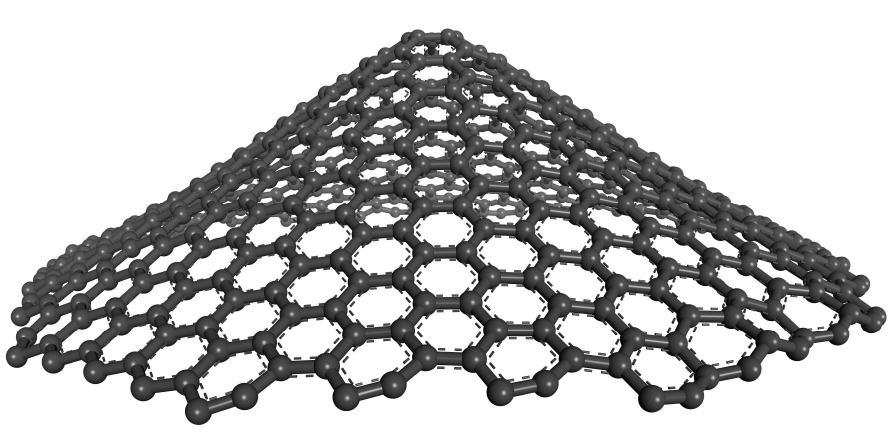}
	\caption{Nanocone obtained from graphene Sheet 1 with disclination angle $\Delta\phi=60^{\circ}$.}
	\label{fig2}
\end{figure}

To determine the radius, we measure the distances between edge atoms connected by a straight line and select the largest distance as the structure's diameter. The radius is then approximated as half of this diameter. This value is identified with the radius $R$ of the defect, as presented in equation (\ref{eq34}) for the surface integral.
To construct the nanocone, we cut a sheet, as demonstrated for Sheet 1 (with radius $2R_{1} = 48.765 \AA$), for an angle of $60^{\circ}$, as shown in Figure \ref{fig1}. The choice of this order of magnitude for the radius has no fundamental justification; it merely ensures a manageable size for constructing the structure and performing its computational simulation.
Next, we solder the edges of the sheet. The resulting nanocone is presented in Figure \ref{fig2}.
Before simulating, we adjust all the border atoms to ensure that their valence is complete and no atom is left bonded to only one other atom, i.e., all carbon atoms are in aromatic rings before soldering.

\begin{figure}[htbp]
	\centering
	\includegraphics[width=0.50\textwidth]{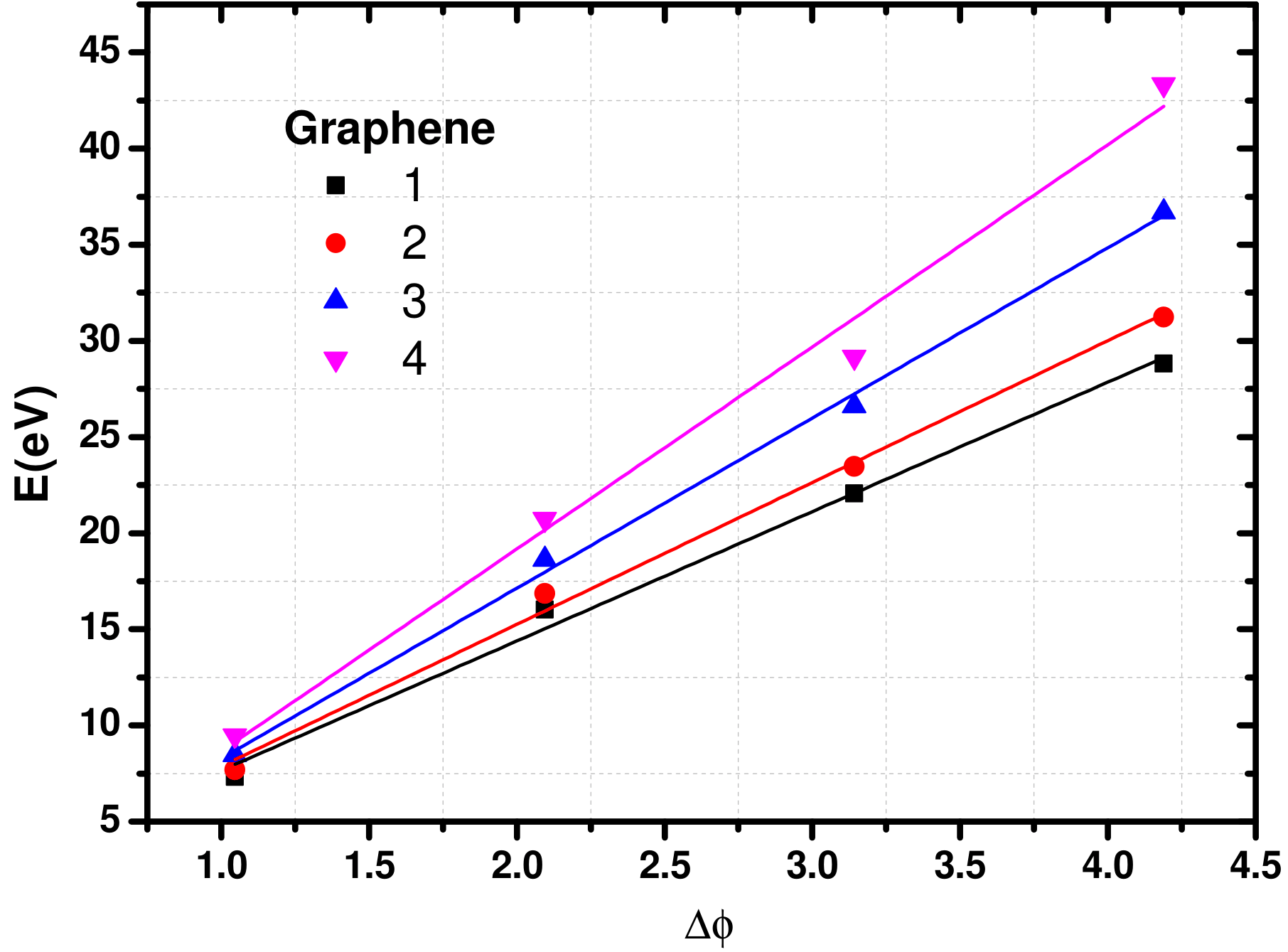}
	\caption{Energy associated with the torsion potential (in $\text{eV}$) for the graphene nanocones as a function of the disclination angle deficit $\Delta\phi$. The dots indicate the data from the simulation, and the lines represent the linear fit.}
	\label{fig3}
\end{figure}

The third step is to optimize the geometry of the structure to find the minimum of the potential (\ref{potential}). The energy associated with the torsion potential can then be obtained from the terms (\ref{torsion}) (not necessarily at the minimum of the potential $V_{torsion}$) that yielded a minimum for $V(R)$. For the nanocone displayed in Figure \ref{fig2}, we obtain an energy of $7.31562\, \text{eV}$. We aim to corroborate a qualitative result, namely the linear behavior of the energy (\ref{eq34}), which we identify as the energy associated with the torsion potential given by the simulation, with respect to the disclination angle. Therefore, we need to consider the same Sheet 1 for different disclination angles, i.e., for $n=2,3,4$ in equation (\ref{angles}). By performing the other three simulations, we obtain the behavior indicated by the black squares in Figure \ref{fig3}.
We can observe an almost linear behavior. By performing a linear fit $E=\alpha \Delta\phi+E_{0}$ of the data, we obtain for Sheet 1
\begin{equation}\label{energy1}
	E_{1} = 6.73327 \, \Delta{\phi}+0.924306
\end{equation}
with a correlation of $0.994$.

To strengthen our results, we considered three additional sheets to mitigate potential coincidences that may arise from analyzing only a single system. In other words, we selected multiple distinct samples (sheets) to demonstrate the universality of the observed behavior. Another motivation for this approach is the absence of a perfect circle: smaller sheets deviate from the ideal circular shape assumed in (\ref{eq34}), while larger ones tend to warp, thereby distorting the nanocones. By examining multiple sheets, we can observe the expected linear behavior both with and without significant deviations from a perfect circle and with or without pronounced nanocone warping. Consequently, we increased the radii as much as feasible while still allowing for the manual application of the Volterra process. The additional sheets are Sheet 2 (with radius $2R_{2}=54.560$ \AA), Sheet 3 (with radius $2R_{3}=68.886$ \AA), and Sheet 4 (with radius $2R_{4}=98.380$ \AA), and performed the same procedure, i.e., four nanocones for each sheet. Thus, a total of 16 simulations were conducted for graphene sheets. The results can also be seen in Figure \ref{fig3}. The linear fits are
\begin{align}
	E_{2} &=  7.377 \,\Delta{\phi} + 0.501\,,\label{energy2}\\
	E_{3} &=  8.854\,\Delta{\phi} - 0.574 \,,\label{energy3}\\
	E_{4} &=  10.516\,\Delta{\phi} - 1.858 \,,\label{energy4}
\end{align}
for Sheets 2, 3, and 4, respectively. The correlations for the linear fits are $0.996$, $0.998$, and $0.991$, respectively. The core result is not the value of the linear coefficients, but the strong linear correlation observed. All sheets exhibit correlations close to unity. It is expected that the linear coefficients will diverge, as the different sizes will influence the uncontrolled effects of circular deviation and warping of the structures.

TEGR predicts an energy independent of the radius, characterized by a universal constant (specific to the material) multiplied by the disclination angle. This universal constant likely depends on an unknown coupling constant $\kappa$ and the thickness of the sheet $L$. Since we do not know the exact value of constant $\kappa$, we can only rely on qualitative corroboration for equation (\ref{eq34}). Further discussion on the nature of this constant in graphene and BN will be presented in Section \ref{sec.5}.

The findings illustrated in Figure \ref{fig3} reveal a clear linear relationship between energy associated with the torsion potential and the disclination angle, aligning closely with the predictions of equation (\ref{eq34}). This alignment leads us to view these results as providing qualitative support for our theoretical framework, and as a qualitative validation of TEGR.

Upon analyzing equations (\ref{energy2})--(\ref{energy4}), we notice a non-zero energy associated with the torsion potential when the flat sheet should ideally have zero energy according to equation (\ref{eq34}). We have verified that all graphene flat sheets exhibit zero energy associated with the torsion potential in the simulation. Therefore, considering that the constant terms in equations (\ref{energy2})--(\ref{energy4}) are one order of magnitude smaller than the linear coefficient, we disregard these constants as errors.

The potential causes for the deviation of the linear factor, particularly its dependence on the radius of the original sheet, will be discussed in Section \ref{sec.5}.

%%%%%%%%%%%%%%%%%%%%%%%%%%%%%%%%%%%%%%%%%%%%%%
\subsection{Boron nitride nanocones} \label{sec.4.2}

The significance of the results observed in the graphene nanocone might initially seem specific to graphene alone. However, we posit that the torsional energy expression derived from TEGR is a universal formulation applicable not only to graphene sheets but also to other materials, and even spacetime. In order to improve this assertion, we conducted a similar analysis on another material.

We generated boron nitride (BN) sheets, known for their structural and thermal properties similar to graphene \cite{aldalbahi2015variations}. Boron nitride possesses insulating properties and exhibits a white coloration. Unlike graphene, boron nitride sheets maintain strength with increasing layers and boost a high Young's modulus ($0.865 TPa$) \cite{falin2017mechanical}. These characteristics make boron nitride suitable for forming stable nanocones.

The steps involved in constructing nanocones from BN sheets mirror those for graphene and will not be elaborated on here; readers can refer to the previous subsection for details.

\begin{figure}[htbp]
    \centering
    \includegraphics[width=0.50\textwidth]{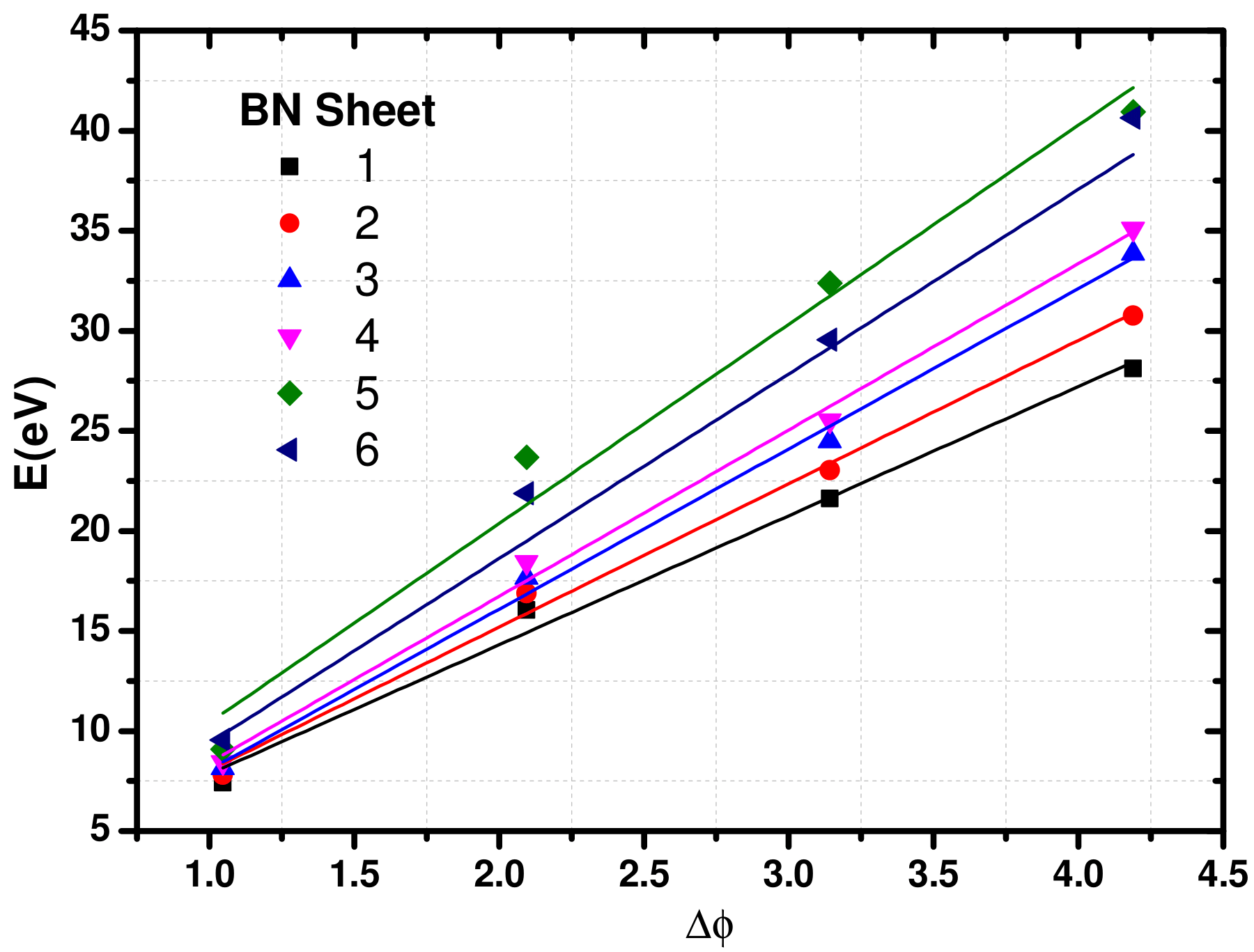}
    \caption{Energy associated with the torsion potential (in $eV$) for the BN nanocones as a function of the disclination angle $\Delta\phi$. The dots represent simulation data, and the lines depict the linear fit.}
    \label{fig4}
\end{figure}

For the BN nanocones, we selected six initial sheets of varying radii: Sheet BN 1 (radius $2R_{1BN}=46.765$\AA), Sheet BN 2 (radius $2R_{2BN}=54.560$\AA), Sheet BN 3 (radius $2R_{3BN}=63.354$\AA), Sheet BN 4 (radius $2R_{4BN}=70.148$\AA), Sheet BN 5 (radius $2R_{5BN}=83.138$\AA), and Sheet BN 6 (radius $2R_{6BN}=98.727$\AA). The outcomes of the 24 simulations are presented in Figure \ref{fig4}.
Once again, the energy associated with the torsion potential exhibits an almost linear growth with the disclination angle across all six structures. This linear trend is quantified through linear regression on Figure \ref{fig4}, yielding the equations
\begin{align}
    E_{1BN} &= 6.456\, \Delta\phi + 1.392\,,\label{energyBN1} \\
    E_{2BN} &= 7.166\, \Delta\phi + 0.858\,, \\
    E_{3BN} &= 8.017\, \Delta\phi + 0.051\,, \\
    E_{4BN} &= 8.314\, \Delta\phi + 0.100\,, \\
    E_{5BN} &= 9.954\, \Delta\phi + 0.461\,, \\
    E_{6BN} &= 9.640\, \Delta\phi + 0.164\,,\label{energyBN6} 
\end{align}
along with their respective correlations: $0.992$, $0.995$, $0.996$, $0.996$, $0.981$, and $0.993$ for Sheets 1 to 6, respectively. All correlations are close to unity.

In equations (\ref{energyBN1}-\ref{energyBN6}), we also note a small constant term, albeit with a significantly smaller magnitude. Furthermore, we observe a linear increase in the coefficient $\alpha$ with the radius of the structure but, like for graphene, not in a linear manner. Sheet BN 1 had a radius less than half of Sheet BN 6, yet this difference did not affect the ratio of the linear coefficients in the same magnitude.

Ergo, we observe that the linear behavior of the energy, associated with the torsion potential, with disclination angles does not seem to be confined to a single material.

%%%%%%%%%%%%%%%%%%%%%%%%%%%%%%%%%%%%%%%%%%%%%%
\section{Final considerations} \label{sec.5}
%%%%%%%%%%%%%%%%%%%%%%%%%%%%%%%%%%%%%%%%%%%%%%

In this article, we delve into the analogies between material physics and the structure of spacetime. Several analogies between condensed matter physics and gravitation can be found in the literature \cite{kraus2019unexpected,fagnocchi2008analog}. We were able to construct a topological defect in spacetime and an equivalent one on graphene and BN monolayers. From the energy definition of TEGR, we predicted a linear response to the disclination angle in the total energy of spacetime. By identifying this total energy with the geometrical (torsional) energy of the nanocones, we found that the definitions of TEGR can be used to evaluate the torsional energy of the material if the constant, analogous to the gravitational coupling one, of the material is determined. We conjecture that these results are not coincidental but part of a greater result, i.e., the energy expression (\ref{eq34}) is valid not only for sheets of any material but for all material structures. This conjecture, together with the known analogies between condensed matter physics and gravity, is a strong indication that the description of gravity by TEGR is more precise than GR. Since the dynamics of TEGR are equivalent to those of GR, all dynamical analogies between GR and material science are valid in TEGR. However, the absence of a consistent definition of gravitational energy in GR makes the results presented here impossible to obtain in Einstein's theory. Therefore, the results of this article strongly favor TEGR. Given that graphene nanocones are not merely theoretical structures, since they have already been synthesized \cite{experimental}, further experimental investigations could verify the present theoretical results.

The main results presented in Figures \ref{fig3} and \ref{fig4} show small deviations from a perfectly linear behavior independent of the size of the structure. It is worth noting that the UFF theory presented in section \ref{sec.3} is a good approximation compared to experimental results, i.e., it contains several classical approximations to a quantum problem. Without these approximations the evaluation would be virtually impossible for huge molecular systems like the nanocones. Besides, it is a numerical solution, and small deviations are to be expected. Hence, similar to experimental data, simulation data contains an uncertainty that cannot be removed. Moreover, we have four other causes for the errors.

\begin{figure}[htbp]
	\centering
		\includegraphics[width=0.40\textwidth]{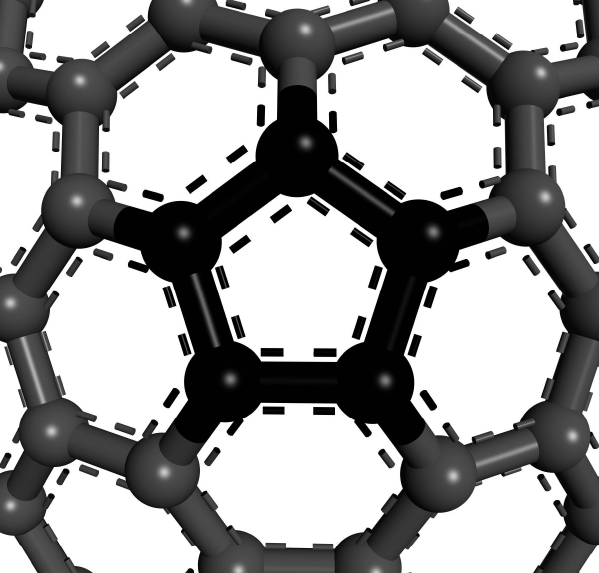}
	\caption{Disclination core of the nanocone for Sheet 1 with $\Delta\phi=60^{\circ}$.}
	\label{fig5}
\end{figure}

\begin{figure}[htbp]
	\centering
		\includegraphics[width=0.40\textwidth]{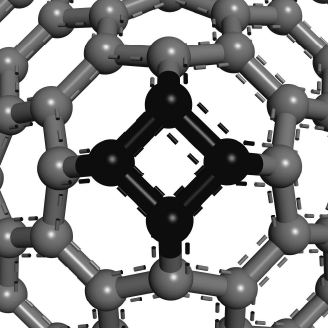}
	\caption{Disclination core of the nanocone for Sheet 1 with $\Delta\phi=120^{\circ}$.}
	\label{fig6}
\end{figure}

\begin{figure}[htbp]
	\centering
		\includegraphics[width=0.40\textwidth]{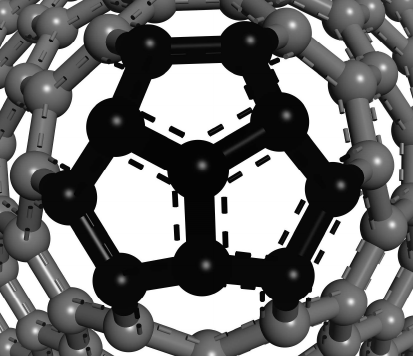}
	\caption{Disclination core of the nanocone for Sheet 1 with $\Delta\phi=180^{\circ}$.}
	\label{fig7}
\end{figure}

\begin{figure}[htbp]
	\centering
		\includegraphics[width=0.40\textwidth]{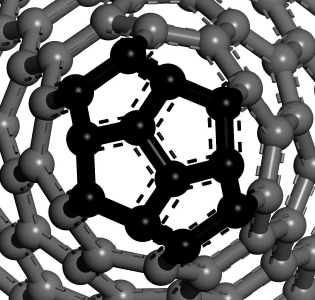}
	\caption{Disclination core of the nanocone for Sheet 1 with $\Delta\phi=240^{\circ}$.}
	\label{fig8}
\end{figure}

First, the model constructed in section \ref{sec.2.1} considers a circular disclination core with radius $r=a$. In the limit $a\rightarrow 0$, we have a disclination core along the $z$ axis only, i.e., a cosmic string. In the real case of the structures, the hexagonal pattern of its atoms makes it impossible to have a punctual disclination core, even a circular one with a small radius. In order to mitigate these effects, we considered structures with an intermediate size, avoiding structures with a small radius. For structures with a diameter around eight hexagons, the correlation to the linear behavior was smaller. The size of the structure does not carry out to the core; thus, with intermediate structures, the contribution of these regions is smaller in comparison. One may think that this contribution is the same for the same sheet, but in reality, the configuration of the disclination core changes with the disclination angle. Let us consider Sheet 1, for example. The core(peak) of the nanocone for $\Delta\phi = 60^{\circ}$ can be seen in Figure \ref{fig5}.
Here, we observe a pentagon in the disclination core. When $\Delta\phi = 120^{\circ}$, a square appears on top, as illustrated in Figure \ref{fig6}.
For $\Delta\phi = 180^{\circ}$ and $\Delta\phi = 240^{\circ}$, as shown in Figures \ref{fig7} and \ref{fig8} respectively, we observe three and four pentagons forming the core.
The border of the disclination core not only changes with the angle but also its size. These distinct cores are necessary to maintain local flatness and minimize the second source of error.

Secondly, we evaluate the energy of a global defect, where spacetime is locally flat. Therefore, the nanocone's sides must not bend to avoid additional, untraceable sources of energy. While the choices (\ref{angles}) minimize local bending, it does not eliminate it. For very large structures, bending occurs near the borders, deviating from metric (\ref{eq24}). This source of error restricted us from choosing very large structures due to bending and very small structures due to the disclination core deviation from circular symmetry.

Thirdly, in our spacetime model, we considered a continuous and infinite spacetime, which is not the case for the sheets. They have a finite radius, resulting in a border where atomic bonds differ from the rest of the structure. These loose atoms increase with the initial sheet's size, potentially affecting the structure's properties unpredictably, and causing the energy to vary with the radius. Also, since the sheets are not perfect circles, the integration contour in (\ref{eq34}) cannot be a perfect circle with radius $r=R$ in reality, as can be seen in Figure \ref{fig2}. As the approximation to a circle improves with increasing size, we expect this effect to vary with the initial sheet's size, influencing the linear coefficient between structures. While we attempted to normalize the energy with the area or perimeter of the initial sheet, we found no satisfactory results. Normalizing with the square root of the radius seemed to approximate the linear coefficients, but we couldn't explain this and left it unaccounted for.

Finally, the UFF is non-relativistic. As a result, the UFF cannot accurately represent materials under relativistic dynamics without introducing errors. This limitation is inherent in its design, as it does not account for relativistic effects.

%%%%%%%%%
Given the significance of the qualitative results, we are prompted to ask: how can we quantitatively validate expression (\ref{eq34})? This validation hinges on the nature of the coupling constant $\kappa$. If we interpret spacetime as a material, this constant is associated with some of its internal properties. In SI units, $c\,\kappa$ has the dimension of force. To associate it with the sheet, we must find an internal characteristic of the material represented by $c\,\kappa$. The Young modulus of the material could partially express this constant, but it yields a magnitude order distinct from the simulation results. Another possibility is the interatomic force between the atoms in the structure, which already has the right dimension. Finding this force is challenging, but we can evaluate the bound energy between the atoms, i.e., the atom-atom electronic interactions. With the bound energy, we can estimate the force by dividing it by the interatomic mean distance. This yields a rough estimate for the force, albeit with the correct magnitude order.

In order to conduct the simulation, we employed another molecular theory, Density Functional Theory (DFT). DFT is an \textit{ab initio} theory that allows for the study of molecular dynamics by describing the electrons as an electronic density field. Unlike the UFF, it is a quantum approach to the problem. Since the theoretical background of this theory is extensive and these simulations are not the main results of our article, we refer the reader to the literature for detailed information \cite{parr1983density,gross2013density}.

Initially, we considered a section with 400 atoms in an infinite graphene sheet. The simulation provided a binding energy of $-3176.27\,eV$. In a graphene sheet, each carbon atom forms three aromatic bonds with adjacent carbon atoms, but each bond is shared between two atoms. Therefore, the total number of bonds can be calculated using the formula
\begin{equation}
	\text{Total number of bonds} = \frac{3 \times \text{Number of atoms}}{2}\,.
\end{equation}
For a sheet with 400 carbon atoms, we have
\begin{equation}
\text{Total number of bonds} = \frac{3 \times 400}{2} = 600\,.
\end{equation}
Thus, there are 600 covalent aromatic bonds in a graphene sheet with 400 carbon atoms. Dividing the binding energy by 600 yields the energy per bond
\begin{equation}
	\epsilon_{C}=-5.294\,eV\,.
\end{equation}
Given the distance $d=1.42 \AA$ between each pair of atoms, the estimated force between atoms is
\begin{equation}
	F_{C}=-\epsilon/d=3.733\,eV/\AA\,.
\end{equation}
From (\ref{eq34}), the predicted linear coefficient $\alpha$ is
\begin{equation}\label{coefC}
	\alpha_{C}=2c \kappa L= 10.452\,eV\,,
\end{equation}
where $L=140\,pm$ (approximate diameter of a carbon atom) is the thickness of the sheet. We can compare (\ref{coefC}) with the results of the linear fit (\ref{energy1}-\ref{energy4}). This rough calculation provides a very good approximation with the correct magnitude order and even a close numerical match for Sheet 4, where we obtained $\alpha=10.516\,eV$. Notably, larger sheets yielded more accurate results.

Secondly, for boron nitride (BN), a similar analysis yields an energy of
\begin{equation}
	\epsilon_{BN}=-4.700\,eV\,.
\end{equation}
In this case, the distance between atoms is $1.45\,\AA$, and the diameter of boron (the larger atom) is $150\,pm$. Hence, we obtain
\begin{equation}
	F_{BN}=3.232\,eV/\AA\,,
\end{equation}
and
\begin{equation}\label{coefBN}
	\alpha_{BN}=2c \kappa L= 9.965\,eV\,.
\end{equation}
Again, we observe the same magnitude order and a very good numerical approximation for Sheet BN 6, where $\alpha=9.226\,eV$. Additionally, we expect lower general values for the coefficient in the BN sheet, as the force between atoms is slightly smaller, reflected in the smaller values in equations (\ref{energyBN1}-\ref{energyBN6}).

If the preceding hypothesis is accurate, as substantiated by these findings, and we conceptualize spacetime as a cellular structure comprised of minuscule particles, we can deduce that the force between these particles corresponds to the coupling constant. Expressed in SI units, this equation results in $c\,\kappa=\frac{c^{4}}{16\pi G}=2.41596\times10^{42}N$. The immense magnitude of this force indicates that spacetime possesses an extremely high resistance to deformation (about $10^{30}$ times greater than that of a one-square-meter graphene sheet). Consequently, gravity, which arises from the deformation of spacetime, is a remarkably feeble phenomena. If the force of deformation is insufficient, meaning that smaller amounts of energy cannot cause considerable deformation of spacetime. This aligns with our understanding of gravity.

\textbf{Acknowledgements:} The authors are grateful to Professor J. W. Maluf for his invaluable assistance in revising the manuscript and for providing insightful ideas that greatly contributed to this work, and the author B.C.C. Carneiro is grateful to R. C. Puraca for assistance in Figure \ref{fig0}.

%\textbf{Conflict of Interest:} The authors declare that there is no conflict of interest.

%\textbf{Data Availability Statement:} Data sharing not applicable – no new data generated, or the article describes entirely theoretical research

%%%%%%%%%%%%%%%%%%%%%%%%%%%%%%%%%%%%%%%%%%%%%%%%%%%%%%

%\bibliographystyle{unsrt}
%\bibliography{bibitex}

\begin{thebibliography}{10}

\bibitem{barcelo}
C.~Barcelo, S.~Liberati, and M.~Visser.
\newblock Analogue gravity.
\newblock {\em Living reviews in relativity}, 14:1--159, 2011.

\bibitem{Unruh}
W.~G. Unruh.
\newblock Experimental black-hole evaporation?
\newblock {\em Physical Review Letters}, 46(21):1351, 1981.

\bibitem{Unruh1}
W.~G. Unruh.
\newblock Has hawking radiation been measured?
\newblock {\em Foundations of Physics}, 44:532--545, 2014.

\bibitem{li2023robophysical}
S.~Li, H.~N. Gynai, S.~W. Tarr, E.~Alicea-Mu{\~n}oz, P.~Laguna, G.~Li, and
  D.~I. Goldman.
\newblock A robophysical model of spacetime dynamics.
\newblock {\em Scientific Reports}, 13(1):21589, 2023.

\bibitem{maluf}
J.~W. Maluf.
\newblock The teleparallel equivalent of general relativity.
\newblock {\em Annalen der Physik}, 525(5):339--357, 2013.

\bibitem{annalen}
J.~W. Maluf, F.~L. Carneiro, S.~C. Ulhoa and J.~F. da~Rocha-Neto.
\newblock Tetrad Fields, Reference Frames, and the Gravitational Energy-Momentum in the Teleparallel Equivalent of General Relativity
\newblock {\em Annalen der Physik}, 535(12):2300241, 2023.

\bibitem{grafeno}
A.~K. Geim.
\newblock Graphene: status and prospects.
\newblock {\em science}, 324(5934):1530--1534, 2009.

\bibitem{referee1}
Capozziello, S., Pincak, R., Saridakis, E. N.
\newblock Constructing superconductors by graphene Chern-Simons wormholes.
\newblock {\em Annals of Physics}, 390, 303-333, 2018.

\bibitem{referee2}
Sepehri, A., Pincak, R., Bamba, K., Capozziello, S., Saridakis, E. N.
\newblock Current density and conductivity through modified gravity in the graphene with defects.
\newblock {\em International Journal of Modern Physics D}, 26(09), 1750094, 2017.

\bibitem{zubkov2015emergent}
M.~A Zubkov and G.~E. Volovik.
\newblock Emergent gravity in graphene.
\newblock In {\em Journal of Physics: Conference Series}, volume 607, page
  012020. IOP Publishing, 2015.

\bibitem{katanaev2005geometric}
M.~O. Katanaev.
\newblock Geometric theory of defects.
\newblock {\em Physics-Uspekhi}, 48(7):675, 2005.

\bibitem{katanaev2020point}
M.~O. Katanaev and B.~O. Volkov.
\newblock Point disclinations in the chern--simons geometric theory of defects.
\newblock {\em Modern Physics Letters B}, 34(supp01):2150012, 2020.

\bibitem{holz1988geometry}
A.~Holz.
\newblock Geometry and action of arrays of disclinations in crystals and
  relation to (2+ 1)-dimensional gravitation.
\newblock {\em Classical and Quantum Gravity}, 5(9):1259, 1988.

\bibitem{katanaev2021disclinations}
M.~O. Katanaev.
\newblock Disclinations in the geometric theory of defects.
\newblock {\em Proceedings of the Steklov Institute of Mathematics},
  313(1):78--98, 2021.

\bibitem{katanaev2003wedge}
M.~O. Katanaev.
\newblock Wedge dislocation in the geometric theory of defects.
\newblock {\em Theoretical and mathematical physics}, 135:733--744, 2003.

\bibitem{katanaev2023combined}
M.~O. Katanaev and A.~V. Mark.
\newblock Combined screw and wedge dislocations.
\newblock {\em Universe}, 9(12):500, 2023.

\bibitem{katanaev2016rotational}
M.~O. Katanaev.
\newblock Rotational elastic waves in a cylindrical waveguide with wedge
  dislocation.
\newblock {\em Journal of Physics A: Mathematical and Theoretical},
  49(8):085202, 2016.

\bibitem{de1998geodesics}
F.~De~Padua, A .and Parisio-Filho and F.~Moraes.
\newblock Geodesics around line defects in elastic solids.
\newblock {\em Physics Letters A}, 238(2-3):153--158, 1998.

\bibitem{charlier2001electronic}
J.~C. Charlier and G.~M. Rignanese.
\newblock Electronic structure of carbon nanocones.
\newblock {\em Physical Review Letters}, 86(26):5970, 2001.

\bibitem{adhikari2012stabilities}
K.~Adhikari and A.~K. Ray.
\newblock Stabilities of silicon carbide nanocones: a nanocluster-based study.
\newblock {\em Journal of Nanoparticle Research}, 14:1--14, 2012.

\bibitem{ardeshana2017approach}
B.~Ardeshana, U.~Jani, A.~Patel, and A.~Joshi.
\newblock An approach to modelling and simulation of single-walled carbon
  nanocones for sensing applications.
\newblock {\em AIMS Materials Science}, 4(4):1010--1028, 2017.

\bibitem{experimental}
K.~Shoyama and F.~Wu{\"u}rthner.
\newblock Synthesis of a carbon nanocone by cascade annulation.
\newblock {\em Journal of the American Chemical Society}, 141(33):13008--13012,
  2019.

\bibitem{moller1964conservation}
C.~Moller.
\newblock Conservation laws in the tetrad theory of gravitation.
\newblock In {\em Proceedings of the Conference on Theory of Gravitation,
  Warszawa and Jablonna, 1962}. Gauthier-Villars, PWN-Polish Scientific
  Publishers, 1964.

\bibitem{puntigam1997volterra}
R.~A. Puntigam and H.~H. Soleng.
\newblock Volterra distortions, spinning strings, and cosmic defects.
\newblock {\em Classical and Quantum Gravity}, 14(5):1129, 1997.

\bibitem{maluf1997gravitational}
J.~W. Maluf and A.~Kneip.
\newblock Gravitational energy of conical defects.
\newblock {\em Journal of Mathematical Physics}, 38(1):458--465, 1997.

\bibitem{maluf2001space}
J.~W. Maluf and A.~Goya.
\newblock Space--time defects and teleparallelism.
\newblock {\em Classical and Quantum Gravity}, 18(23):5143, 2001.

\bibitem{carneiro2020quantization}
F.~L. Carneiro, S.~C. Ulhoa, J.~F. da~Rocha-Neto, and J.~W. Maluf.
\newblock On the quantization of burgers vector and gravitational energy in the
  space-time of a conical defect.
\newblock {\em The European Physical Journal C}, 80:1--9, 2020.

\bibitem{bretonnet2017basics}
J.~L. Bretonnet.
\newblock Basics of the density functional theory.
\newblock {\em AIMS Materials Science}, 4(6):1372--1405, 2017.

\bibitem{viana2004quantitative}
J.~M.~S. Viana.
\newblock Quantitative genetics theory for non-inbred populations in linkage
  disequilibrium.
\newblock {\em Genetics and Molecular Biology}, 27:594--601, 2004.

\bibitem{rappe1992uff}
A.~K. Rapp{\'e}, C.~J. Casewit, K.~S. Colwell, W.~A. Goddard~III, and W.~M.
  Skiff.
\newblock Uff, a full periodic table force field for molecular mechanics and
  molecular dynamics simulations.
\newblock {\em Journal of the American chemical society}, 114(25):10024--10035,
  1992.

\bibitem{jaillet2017uff}
L.~Jaillet, S.~Artemova, and S.~Redon.
\newblock Im-uff: Extending the universal force field for interactive molecular
  modeling.
\newblock {\em Journal of Molecular Graphics and Modelling}, 77:350--362, 2017.

\bibitem{mackerell2004empirical}
A.~D. MacKerell~Jr.
\newblock Empirical force fields for biological macromolecules: overview and
  issues.
\newblock {\em Journal of computational chemistry}, 25(13):1584--1604, 2004.

\bibitem{bradley1972ap}
C.~J. Bradley.
\newblock Ap cracknell the mathematical theory of symmetry in solids,
  clarendon, 1972.

%\bibitem{tajkhorshid2002control}
%E.~Tajkhorshid, P.~Nollert, M.~{\O}. Jensen, L.~J.~W. Miercke, J.~O'Connell,
 % R.~M. Stroud, and K.~Schulten.
%\newblock Control of the selectivity of the aquaporin water channel family by
  %global orientational tuning.
%\newblock {\em Science}, 296(5567):525--530, 2002.

\bibitem{leach2001molecular}
A.~R. Leach.
\newblock {\em Molecular modelling: principles and applications}.
\newblock Pearson education, 2001.

\bibitem{mayne2013rapid}
C.~G Mayne, J.~Saam, K.~Schulten, E.~Tajkhorshid, and J.~C. Gumbart.
\newblock Rapid parameterization of small molecules using the force field
  toolkit.
\newblock {\em Journal of computational chemistry}, 34(32):2757--2770, 2013.

\bibitem{vanommeslaeghe2014molecular}
K.~Vanommeslaeghe, O.~Guvench, et~al.
\newblock Molecular mechanics.
\newblock {\em Current pharmaceutical design}, 20(20):3281--3292, 2014.

\bibitem{tute1995drug}
M.~S. Tute.
\newblock Drug design: The present and the future.
\newblock In {\em Advances in Drug Research}, volume~26, pages 45--142.
  Elsevier, 1995.

\bibitem{alfieri2011structural}
G.~Alfieri and T.~Kimoto.
\newblock Structural stability and electronic properties of sic nanocones:
  first-principles calculations and symmetry considerations.
\newblock {\em Applied Physics Letters}, 98(12), 2011.

\bibitem{aldalbahi2015variations}
A.~Aldalbahi, A.~F. Zhou, and P.~Feng.
\newblock Variations in the crystal structures and electrical properties of
  single-crystal boron nitride nanosheets.
\newblock {\em Scientific reports}, 5(1):16703, 2015.

\bibitem{falin2017mechanical}
A.~Falin, Q.~Cai, E.~J.~G. Santos, D.~Scullion, D.~Qian, R.~Zhang, Z.~Yang,
  S.~Huang, K.~Watanabe, T.~Taniguchi, et~al.
\newblock Mechanical properties of atomically thin boron nitride and the role
  of interlayer interactions.
\newblock {\em Nature communications}, 8(1):15815, 2017.

\bibitem{kraus2019unexpected}
P.~Kraus.
\newblock The unexpected duality of gravitational and condensed-matter physics.
\newblock {\em Physics Today}, 72(4):56--57, 2019.

\bibitem{fagnocchi2008analog}
S.~Fagnocchi.
\newblock Analog models beyond kinematics.
\newblock In {\em The Logic Of Nature, Complexity And New Physics: From
  Quark-Gluon Plasma to Superstrings, Quantum Gravity and Beyond}, pages
  555--564. World Scientific, 2008.

\bibitem{parr1983density}
R.~G. Parr.
\newblock Density functional theory.
\newblock {\em Annual Review of Physical Chemistry}, 34(1):631--656, 1983.

\bibitem{gross2013density}
E.~K.~U. Gross and R.~M. Dreizler.
\newblock {\em Density functional theory}, volume 337.
\newblock Springer Science \& Business Media, 2013.

\end{thebibliography}

\end{document}